\documentclass[10pt,conference]{IEEEtran}
\IEEEoverridecommandlockouts

\usepackage{amsmath,amssymb,amsfonts}
\usepackage{algorithmic}
\usepackage{graphicx}
\usepackage{textcomp}
\usepackage{xcolor}
\usepackage[
style=ieee,
maxnames=1,
minnames=1,
]{biblatex}
\usepackage{hyperref}
\usepackage{cleveref}
\usepackage{siunitx}
\usepackage[absolute,overlay]{textpos} 
\usepackage{tcolorbox} 
\usepackage{caption}

\usepackage{todonotes}
\usepackage[inline]{enumitem}

\graphicspath{{./images}}

\usepackage{tikz}
\usetikzlibrary{automata,shapes.geometric, arrows.meta, positioning, calc}

\usepackage{pbox}
\title{
{
    A2P: A Scalable OFDMA Polling Algorithm for Time-Sensitive Wi-Fi Networks
{\footnotesize \textsuperscript{}
}}}

\author{\IEEEauthorblockN{Douglas Dziedzorm Agbeve\IEEEauthorrefmark{1}, Andrey
    Belogaev\IEEEauthorrefmark{1}, Wim Sandra\IEEEauthorrefmark{2}, Carl
Lylon\IEEEauthorrefmark{2}, Jeroen Famaey\IEEEauthorrefmark{1}}
\IEEEauthorblockA{
    \textit{\IEEEauthorrefmark{1}IDLab, University of Antwerp – IMEC,
    Belgium}\\
    \textit{\IEEEauthorrefmark{2}Televic Group, Belgium} \\
\{douglas.agbeve, andrei.belogaev, jeroen.famaey\}@uantwerpen.be \\
\{w.sandra, c.lylon\}@televic.com}}

\begin{document}
\maketitle
\begin{textblock*}{\textwidth}(1.6cm, 25.7cm) 
    \centering
    \begin{tcolorbox}[colframe=black,
        colback=white,
        boxrule=0.4pt,
        width=0.98\textwidth,
         ] 

        \small \textcopyright~2025 IEEE. Personal use of this material is
        permitted. Permission from IEEE must be obtained for all other uses, in
        any current or future media, including reprinting/republishing this
        material for advertising or promotional purposes, creating new
        collective works, for resale or redistribution to servers or lists, or
        reuse of any copyrighted component of this work in other works.
    \end{tcolorbox}
\end{textblock*}
\begin{abstract}
Over the years, advancements such as increased bandwidth, new modulation
and coding schemes, frame aggregation, and the use of multiple antennas have
been employed to enhance Wi-Fi performance. Nonetheless, as network density and
the demand for low-latency applications increases, contention delays and
retransmissions have become obstacles to efficient Wi-Fi deployment
in modern scenarios. The introduction of Orthogonal Frequency-Division Multiple
Access (OFDMA) in the IEEE 802.11 standard allows simultaneous transmissions to
and from multiple users within the same transmission opportunity, thereby
reducing the contention.
However, the AP must efficiently manage the resource allocation, particularly in
uplink scenarios where it lacks prior knowledge of users' buffer statuses, thus
making polling a critical bottleneck in networks with a high number of users
with sporadic traffic pattern. This paper addresses the polling problem and
introduces the A2P algorithm, designed to enable scalable and efficient polling
in high-density OFDMA-based time sensitive Wi-Fi networks.  Simulation results
show that A2P outperforms the alternative schemes by maintaining significantly
lower delay and packet loss in dense time-sensitive teleconferencing scenario.
\end{abstract}


\vspace{-0.5cm}
\section{Introduction}

Historically, Wi-Fi technology standardized as IEEE 802.11 has always relied
solely on random channel access procedures, Distributed Coordination Function
(DCF) or Enhanced Distributed Channel Access (EDCA). One of the main advantages
of such schemes is that stations can transmit data on demand, without the need
for the access point (AP) to coordinate the schedule. Throughout the development
of Wi-Fi, different approaches were applied to increase its performance in terms
of the data rate, such as bandwidth increase, advanced modulation and coding
schemes,  frame aggregation, and multiple transmitting and receiving antennas.
However, with increasing network density and application demands for low
latency, the overhead corresponding to the contention delays and retransmissions
has become a significant obstacle for applying Wi-Fi to many modern scenarios,
which often encompass time sensitive traffic. In contrast, channel orchestration
delays for scheduled access have become lower due to increased data rate. Hence,
since the introduction of the IEEE 802.11ax standard~\cite{80211ax}, OFDMA is
available as an alternative channel access method in Wi-Fi.

Although OFDMA is relatively new to Wi-Fi, it has long been used for wireless
communication in cellular systems, such as 4G, 5G, and beyond. OFDMA allows
simultaneous transmissions to/from multiple users in the same transmission
opportunity, thus reducing the contention between them. To achieve this, the
bandwidth is divided into multiple subcarriers, called Resource Units (RUs),
that can be allocated by the AP to different users. Hence, the AP needs to know
the buffer status of those users to avoid allocating resources to users with
empty buffers. In downlink, the AP knows how many packets/bytes it needs to send
to each user, but in uplink it has to poll the users to obtain buffer status
reports (BSRs). When the number of users increases, polling becomes the
bottleneck, because the AP does not know in advance which users are going to
have data to transmit. In literature, this problem did not gain much attention,
and most of the works are focused on optimizing the scheduling procedure in
order to satisfy the quality of service (QoS)
requirements (see more details in Section~\ref{related_works}). In this paper,
we highlight the polling issue, and propose the A2P algorithm that enables
scalable and efficient polling in Wi-Fi networks with a high number of users
with sporadic traffic.

The rest of the paper is organized as follows. Section~\ref{ofdma_overview}
provides an overview of the main OFDMA features and parameters. In
Section~\ref{related_works}, we analyze the state-of-the-art methods and
approaches to configuring and optimizing OFDMA operation in Wi-Fi. In
Section~\ref{system_model}, we describe the system model. Subsequently,
Section~\ref{algorithm} introduces the proposed A2P polling algorithm. In
Section~\ref{performance_evaluation}, we study the performance of the proposed
algorithm against common alternatives and baselines. Finally, in
Section~\ref{conclusion}, we conclude the paper.

\section{OFDMA Operation in Wi-Fi Networks} \label{ofdma_overview}
\tikzstyle{sta_freq} = [rectangle, draw, align=flush center, very  thick,
                        minimum height = .5cm, minimum width = 1cm, outer sep =
                        0pt]
\tikzstyle{ap_freq} = [rectangle, draw, align=flush center,very thick,
                        minimum height = 2.5cm, minimum width = 1.0cm]
\tikzstyle{line} = [-{latex[length=7mm, width=5mm]}, very thick,
                        draw,color=black!100]
\tikzstyle{sifs} = [{latex[length=7mm, width=5mm]}-{latex[length=7mm,
width=5mm]}, thick, draw,color=black!100]
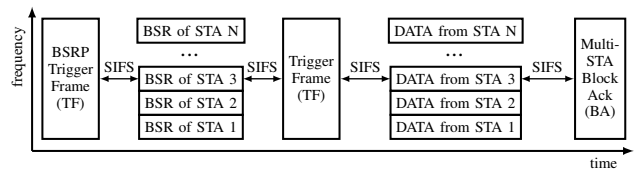
\begin{figure}[tb]
    \centering
        \scalebox{0.64}{
            \centering
            \begin{tikzpicture}[node distance=2.5cm, font=\small, on
                grid, auto, label position=center, align = flush center]
                \node[ap_freq](AP1) {BSRP\\Trigger\\Frame\\(TF)};
                \node[sta_freq, right of = AP1](STA1){BSR of STA 3};
                \node[sta_freq, below of = STA1, node distance=.5cm](STA2){BSR
                    of STA 2};
                \node[sta_freq, below of = STA2, node distance=.5cm](STA3){BSR
                    of STA 1};
                \node[above of = STA1, node distance =
                    .5cm](STA4){\textbf{\dots}};
                \node[sta_freq, above of = STA4, node distance=.5cm](STA5){BSR
                    of STA N};
                \node[ap_freq, right of = STA1](AP2)
                    {Trigger\\Frame\\(TF)};
                 \node[sta_freq, right of = AP2, node distance =
                     3cm](STA6){DATA from  STA 3};
                \node[sta_freq, below of = STA6, node
                    distance=.5cm](STA7){DATA from STA 2};
                \node[sta_freq, below of = STA7, node
                    distance=.5cm](STA8){DATA from STA 1};
                \node[above of = STA6, node distance =
                    .5cm](STA9){\textbf{\dots}};
                \node[sta_freq, above of = STA9, node
                    distance=.5cm](STA10){DATA from STA N};
                \node[ap_freq, right of = STA6, node distance = 3cm](AP3)
                    {Multi-\\STA\\Block\\Ack\\(BA)};

                \path[line](-.8,-1.5) -- node[pos=.65, rotate=90,
                    anchor=south]{frequency}++(0,3);
                \path[line](-.8,-1.5) -- node[pos=0.95,
                    anchor=north]{time}++(12.5,0);
                \path[sifs](AP1)--node[pos=0.5]{SIFS}(STA1);
                \path[sifs](STA1)--node[pos=0.5]{SIFS}(AP2);
                \path[sifs](AP2)--node[pos=0.5]{SIFS}(STA6);
                \path[sifs](STA6)--node[pos=0.5]{SIFS}(AP3);
            \end{tikzpicture}
        }
    \captionsetup{belowskip=-18pt}
    \caption{\footnotesize{UL OFDMA frame exchange sequence}}
    \label{fig:ul_ofdma}
\end{figure}
IEEE~802.11ax, also referred to as Wi-Fi~6, introduced OFDMA to the Wi-Fi
world~\cite{80211ax}. When OFDMA is used, the data transmission for each station
(STA) does not occupy the whole bandwidth. Instead, the bandwidth is divided
into multiple RUs, which are allocated to different STAs. Each RU may consist of
26, 52, 106, 242 (20 MHz), 484 (40 MHz), 996 (80 MHz), 2x996 (160 MHz), and
since IEEE 802.11be (Wi-Fi 7)~\cite{80211be} 4x996 (320 MHz) tones.

The Wi-Fi standard supports both downlink (DL) and uplink (UL) OFDMA
transmissions. Since the focus of this paper is on the polling procedure, that
occurs only in UL OFDMA, further we describe the UL case.

OFDMA operates atop random channel access (EDCA), meaning that, to initiate
multi-user frame exchange, the AP must access the channel by contending with
STAs and other neighboring APs.
Once the AP wins the contention, it can initiate UL OFDMA transmission by
sending a Trigger Frame (TF). This frame contains, among other parameters, a
mapping between the STAs that will transmit and their respective RU allocation.
After a time interval referred to as short inter-frame space (SIFS), the UL
OFDMA transmission follows. Transmissions in one multi-user OFDMA frame must be
synchronized in time. As such, short-length data packets should be padded and/or
aggregated to match the size of the largest packet in this OFDMA frame. A SIFS
after the transmission finishes, the AP replies with a multi-STA Block
Acknowledgement (BA), confirming the reception of the uplink packets from
different users.

To schedule RUs to STAs that have UL data, the AP requires information about the
buffer status of the queues in those STAs. To obtain this information, the AP
polls the STAs with a special TF called buffer status report poll (BSRP) TF.
Similarly to the regular TF, the BSRP TF contains the mapping of different users to
RUs, in which the AP expects the buffer status report (BSR) in response
containing the buffer size in bytes. STAs can provide their BSR explicitly using
a QoS\hyphen Null frame with no payload in response to a BSR Poll (BSRP) TF
matching their Association Identifier (AID), or implicitly through the QoS
control field of any transmitted frame. STAs with non-empty buffers are then
considered for scheduling in the subsequent TF. The complete frame exchange
sequence for the UL OFDMA procedure is depicted in Figure~\ref{fig:ul_ofdma}.

To promote airtime fairness and reduce contention in a densely deployed network,
Wi-Fi introduced the MU EDCA Parameter Set, which contains EDCA parameters
(contention window and arbitrary inter-frame space) that should be used by the
STA participating in OFDMA transmissions. This way, the network can be
configured so that the AP gets more control in orchestrating UL transmissions,
while the STAs themselves compete for the channel less aggressively, or do not
compete for the channel at all. The MU-EDCA Parameter Set is announced by the AP
in beacon frames. Apart from the EDCA parameters, it also contains the validity
timer. Once the timer is expired, the STA reverts to the default EDCA
parameters. The timer is reset at the STA every time successful OFDMA
transmission occurs, i.e., after the successful reception of a BA from the AP.
Thus, the STA maintains the updated parameters as long as it gets resources
scheduled by the AP.

Clearly, explicit solicitation of BSRs from STAs through BSRP TF does not scale
well in scenarios with many STAs associated with the same AP. More specifically,
in scenarios such as large teleconferencing applications, where only a handful
of users are actively talking (i.e., transmitting data), while many others are
following the conversation (i.e., only receiving data), the polling of idle STAs
without data to transmit wastes resources and degrades the overall network
performance. In this paper, we show that inefficient polling leads to violation
of QoS requirements, and propose a solution that outperforms the alternative
schemes in terms of the achieved latency and ratio of packets delivered on
time.

\section{Related Work}
\label{related_works}

Numerous works in literature propose solutions and improvements for OFDMA
performance optimization in Wi-Fi networks. Many papers~\cite{1_Wang, 1_Bankov,
1_Inamullah, 1_Qadri, noh2024joint, tan2024deep} consider resource allocation
algorithms, while BSR collection is not considered as a potential bottleneck for
UL OFDMA performance. In~\cite{1_Wang}, the resource allocation problem was
formulated as an optimization problem and addressed using a sub-optimal
divide-and-conquer recursive algorithm. In~\cite{1_Bankov}, the authors propose
an adaptation of three well-known scheduling algorithms --- MaxRate, Proportional
Fair and Shortest Remaining Processing Time --- to Wi-Fi. In~\cite{1_Inamullah},
the authors propose a scheduling algorithm for deadline-constrained traffic. To
estimate the delay the packet will spend in the queue before the transmission,
the authors apply queuing theory using the buffer statuses reported by the
STAs. In~\cite{1_Qadri}, a heuristic algorithm is designed to prioritize users
based on the reported buffer sizes for each access category (AC). In more recent
papers~\cite{noh2024joint, tan2024deep}, a deep reinforcement learning approach
is applied. For most of the solutions proposed in these papers, the knowledge
of buffer statuses plays a crucial role, but the overhead related to the BSR
collection procedure is left out of scope.

A common approach considered in literature for BSR is to utilize Uplink
OFDMA-based Random Access (UORA) defined in the IEEE 802.11 standard.
Since it is based on random access, collisions are possible. After a collision,
the STA will have to retransmit lost packets, that is why UORA is often
considered as a solution only for BSRs, because they are generally significantly
shorter than data packets. There are multiple papers in literature evaluating
performance of UORA and proposing enhancements. In~\cite{1_Naik,
1_Bhattarai}, the metric BSR delivery rate is introduced.
The performance evaluation shows that BSR delivery is of crucial importance for
the performance of the scheduling algorithm. Besides, the authors propose a resource
allocation algorithm that varies the number of random access RUs depending on
the number of STAs with non-zero BSRs. However, the proposed algorithm will not
allocate any RUs to UORA if the number of STAs with non-zero BSRs is higher or
equal than the number of available RUs, thus significantly favoring the STAs
with already known BSR. In~\cite{1_Avdotin, 2_Avdotin, 3_Avdotin}, the authors
propose different schemes for collision resolution in UORA to reduce latency
related to retransmissions of the lost packets. In this paper, we leave UORA out
of scope for several reasons. First, we argue that UORA sacrifices the part of
available bandwidth leading to its underutilization when there are not many
STAs with new data. Second, UORA is not a mandatory feature in the IEEE 802.11ax
standard. Third, there is no verified and publicly available implementation of
UORA in the available network simulators. However, it will be considered in our
future works.

Recently, a few papers proposed solutions that do not rely on UORA.
In~\cite{1_Schneider}, the authors consider the problem of providing fully
deterministic channel access in Wi-Fi. For that, they propose to completely
disable random access procedures, and let the AP orchestrate all UL
transmissions. Although such an approach is very powerful when the traffic
pattern is known, e.g., when it is periodic, we show in this paper that it will
lead to significant performance degradation in scenarios with many devices with
unpredictable on-off traffic.
In~\cite{shao2024access}, the BSR collection is proposed based on multiple
rounds of BSRP TF and BSR exchanges to collect BSRs from all STAs before
scheduling resources for data transmission. This approach can improve the
fairness by giving more information to the scheduling algorithm before it makes
its decision, but does not solve the problem of BSR collection overhead.
In~\cite{goncalves2023access}, client-side access manipulation is proposed.
According to this manipulation, the STA switches between two channel access
schemes, EDCA and OFDMA, depending on its latest buffer status. In the OFDMA
state, it fully disables EDCA, while in the EDCA state it competes for the
channel with other STAs and the AP. However, according to the standard it is
the AP who controls these two states via the MU EDCA Parameter Set. Moreover,
excluding the AP from the decision process may lead to inconsistent behavior
and resource waste, as the AP does not know the STA's state.

In this paper, we propose a fully standard compliant A2P algorithm that
significantly improves the latency compared to the state of the art. This is
verified in a challenging teleconference scenario with a high number of STAs.

\section{System Model} \label{system_model}
We consider a teleconferencing system, where multiple participants with wireless
microphones and headphones are connected to the server through a single AP. We
assume that the communication channel between the server and the AP is extremely
fast (e.g., fiber connection), hence we consider both as a single entity.
Furthermore, we assume that the microphone and headphones are integrated into a
single tabletop unit, which we call a Station (STA). In this paper, we restrict
the scope of our model to only the bidirectional audio streams,
from microphones to the server and from the server to the headphones. Although
not all microphones are active at the same time, the total number of STAs can be
very high in large-scale teleconferencing settings (e.g., in large debating
halls such as a parliament); easily exceeding \num{100}.

When a user presses it's microphone button to start speaking, the STA generates
an UL constant bit rate (CBR) audio data stream, which stops when the button is
released, resulting in a stochastic on-off data streaming pattern. The streaming
timeline is divided into $X$\unit{\milli\second} windows, and within first $B$
\unit{\milli\second} of the window, the STA can generate the audio packet.
Each packet contains $M$ samples of audio information with $N$
\unit{bit} resolution, and a $W$ \unit{bits} application layer header, resulting
in a packet size of $M \times N+W$ \unit{bits}, and a bitrate of $1000 \times
\left(M \times N+W\right) / X$\unit{bps}.

\begin{figure}[tb]
    \centering
    \vspace{5pt}
    \includegraphics[scale=0.15]{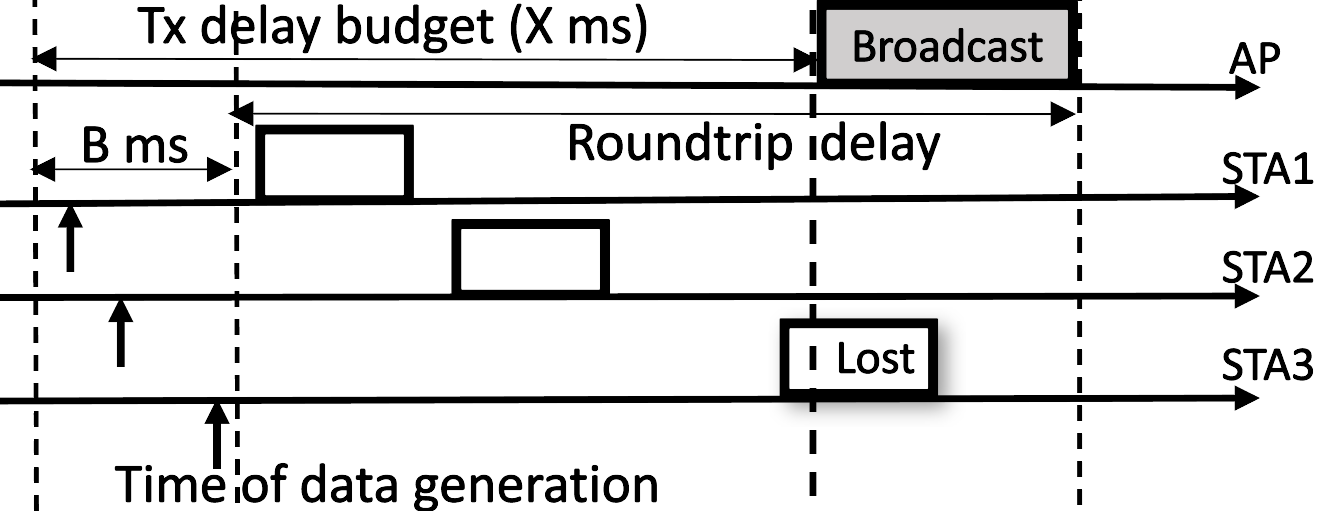}
    \captionsetup{belowskip=-18pt}
    \caption{\footnotesize{Model of system's traffic pattern}}
    \label{fig:system_model_b}
\end{figure}

To avoid overload, the STAs should deliver their packets within the transmission
delay budget ($X$\unit{\milli\second}), defined as the maximum allowable time
for a packet to be received by the AP after its generation. UL packets that
arrive later than $X$\unit{\milli\second} are considered lost, and hence ignored
by the server. For instance, in Figure~\ref{fig:system_model_b}, STA1 and STA2
successfully transmit within the delay budget. In contrast, STA3's packet is
received by the AP after the delay budget, and is, therefore, lost.

After the server receives data packets from all the active STAs or the end of
delay budget is reached, it generates a single DL data packet containing mixed
audio information received from the STAs. This DL packet contains $P$ samples
with resolution of $Q$ \unit{bits}, and an application layer header of $W$
\unit{bits}. The packet size is therefore $P \times Q+W$ \unit{bits}. As it
follows the same packet generation interval $X$\unit{\milli\second} as the STA,
the DL bitrate is $1000 \times  \left(P \times Q+W\right) / X$ \unit{bps}. The
packet is broadcast to all STAs. This way the participants can follow the
ongoing conversation.
The broadcast packet in~Figure~\ref{fig:system_model_b} contains
mixed samples received only from STA1 and STA2. Note that the packet size of the
DL packet is independent of the number of successfully received UL packets.

Since the AP should have access to the channel to orchestrate UL transmissions,
it cannot request it only when it has enqueued DL packets. Hence, it
periodically requests the channel with periodicity $T$ \unit{\milli\second}
called Access Request Interval (ARI). To avoid unnecessary attempts to request
the channel, we count down this interval from the last time the AP obtained
access to the channel.
\section{A2P Polling Algorithm} \label{algorithm}
Our proposal towards a more efficient polling for UL OFDMA involves the
combination of the two main channel access mechanisms in Wi-Fi, EDCA and OFDMA.
The AP maintains a list of active STAs (i.e., STAs that are expected to transmit
data), called the \emph{polling list}. These STAs use only UL OFDMA channel
access, thus reducing the amount of channel time  wasted on EDCA contention. All
other STAs use the EDCA mechanism. To get on the polling list, a STA that wins
access to the channel transmits its initial packet, thus announcing the start of
of data stream. After a successful uplink EDCA data exchange, the AP disables
EDCA on the STA for a period of $Y$ Time Units (TUs) using the MU EDCA Parameter
Set. Meanwhile, STAs that have not reported any data after being polled during
$Y$ TUs are removed from the polling list. If the size of the polling list
exceeds the number of available RUs for the chosen bandwidth, a simple
round-robin method is used to choose which STAs to poll. Additionally, the AP
can request access to the channel only after waiting one ARI from the previous
granted access, provided there is no enqueued downlink frame.

Figure~\ref{fig:sta_flow_chat} illustrates the A2P STA-side procedures as a flow
chart. A STA that gains access to the channel through contention transmits data
and, upon receiving an ACK, disables EDCA for a duration of MU\hyphen EDCA timer
value, i.e., $Y$ TUs. The MU\hyphen EDCA timer is reset to its original value
if, before it counts down to zero, the STA takes part in UL\hyphen OFDMA (i.e.,
successfully transmits frames and receives the BlockAck). Otherwise, EDCA is
re-enabled when the timer reaches zero, and the STA can contend for the channel
again.

The flow chart for the AP side is shown in Figure~\ref{fig:ap_flow_chat}. After
acknowledging the reception of the frames that were transmitted by STAs, the AP
adds the new STAs to the polling list and sets a timer (i.e., $Y$ TUs) for each
STA. When the AP wins channel contention, it decides on either performing DL or
UL frame exchange. For simplicity and fairness between UL and DL, we assume that
these alternate each other, i.e., a DL frame exchange after every UL frame
exchange and vice versa. Since the considered teleconferencing system requires
only a single broadcast DL transmission, this logic does not affect the
performance.
During UL transmission, the AP selects a group of STAs to participate in the
poll-based frame exchange using round-robin scheduling. The maximum number of
selected STAs depends on the number of available RUs for the chosen bandwidth.
To maximize the number of supported simultaneous transmissions, we consider only
the smallest 26-tones RUs. For instance, a \qty{40}{\MHz} channel can serve up to \num{18} STAs. Note that this
assumption is reasonable for audio traffic, because it has a relatively low
bitrate. For heavier traffic it may lead to performance degradation.
Following the UL OFDMA frame exchange, the AP resets the timer of those STAs
that successfully transmitted data.

\tikzstyle{start} = [ellipse, draw, align=flush center, rounded corners, minimum size = .5cm, inner sep=0, outer sep=0]

\tikzstyle{process} = [rectangle, draw, align=flush center, rounded corners, minimum height = 0cm, minimum width = 1cm]

\tikzstyle{decision} = [diamond, draw, align=flush center, rounded corners, inner sep=0, outer sep=0, minimum size = 0cm]

\tikzstyle{arrow_line} = [-{latex[length=7mm, width=5mm]},
                        draw,color=black!100]

\begin{figure}[tb]
    \centering
        \vspace{5pt}
        \scalebox{0.7}{
            \centering
            \begin{tikzpicture} [ font=\small, on
                grid, auto, label position=center, align = flush center]
                \node[start](start){System Setup};
                \node[process, below of=start, node distance = 1.0cm](P1){STA
                        wins access to channel and \\transmits frame(s) in its
                    queue};
                \node[decision, below of = P1, node distance = 2.0cm](D1){ACK\\
                    received\\from \\AP?};
                \node[process, below of = D1, node distance = 2.0cm](P2){ STA
                        disables\\EDCA for \textit{$Y$} TUs};
                \node[decision, below of = P2, node distance =
                    2.2cm](D2){STA\\partook in \\UL OFDMA\\TX?};
                \node[process, left of = D2, node distance =
                    3.5cm](P3){Reset\\EDCA-disabled\\timer (\textit{$Y$} TUs)};
                \node[decision, below of = D2, node distance =3.2cm]
                    (D3){EDCA-disabled\\timer expired?};
                \node[process, left of = D3, node distance = 3.2cm, minimum
                    height= 1cm](P4){Enable EDCA};
                \node[process, left of = D1, node distance = 3.2cm] (P5)
                    {Channel\\contention};

                \path[arrow_line](start)--(P1);
                \path[arrow_line](P1)--(D1);
                \path[arrow_line](D1)--node{yes}(P2);
                \path[arrow_line](D1) -- node[anchor=south,pos=0.5]{no}(P5);
                \path[arrow_line](P2)--(D2);
                \path[arrow_line](D2)--node[anchor=south, pos=0.5]{yes}(P3);
                \path[arrow_line](P3.north) |- (P2.west);
                \path[arrow_line](D2)--node{no}(D3);
                \path[arrow_line](D3)--node[anchor=south, pos=0.5]{yes}(P4);
                \path[arrow_line](P4.west) to + (-1, 0) |- (P5);
                \path[arrow_line](D3.east) to + (1, 0) |- node[near start]{no}
                    (D2.east);
                \path[arrow_line](P5.north)|-(P1.west);
            \end{tikzpicture}
        }
    \captionsetup{belowskip=-12pt}
    \caption{Flow chart of the algorithmic procedure at STA side}
    \label{fig:sta_flow_chat}
\end{figure}
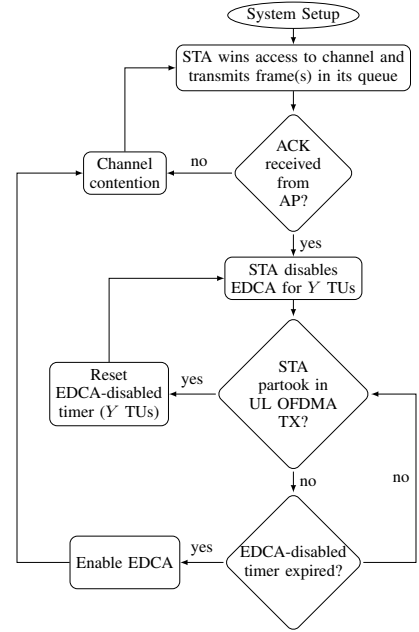
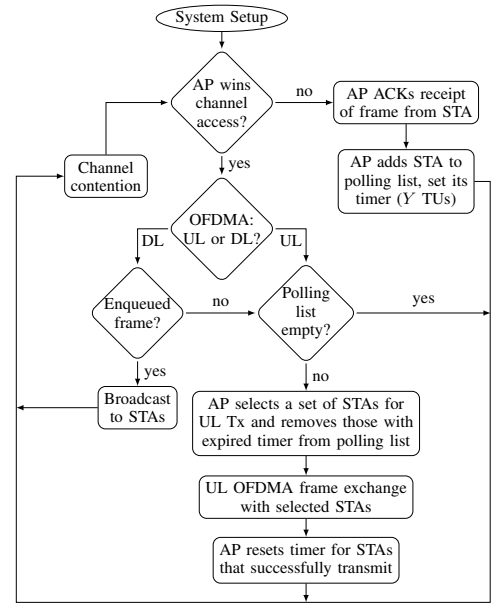
\begin{figure}[tb]
    \centering
        \scalebox{0.7}{
            \centering
            \begin{tikzpicture}[font=\small, on grid, auto, label position=center, align = flush center]
                \node[start](start){System Setup};
                \node[decision, below of = start, node distance = 1.6cm] (D1)
                    {AP wins\\channel \\access?};
                \node[process, right of = D1, node distance = 3.5cm] (P1) {AP
                    ACKs receipt\\of frame from STA};
                \node[process, below of = P1, node distance = 1.5cm] (P2) {AP adds
                    STA to\\polling list, set
                    its\\timer (\textit{$Y$} TUs)};
                \node[decision, below of = D1, node distance = 2.4cm] (D2)
                    {OFDMA:\\UL or DL?};
                \node[decision, left of = D2, below of = D2, node distance =
                    1.6cm] (D3) {Enqueued\\frame?};
                \node[process, below of = D3, node distance = 1.8cm]
                    (P3){Broadcast\\to STAs};
                \node[decision, below of = D2, right of = D2, node distance =
                    1.6cm] (D4) {Polling\\list\\empty?};
                \node[process, below of = D4, node distance = 2.1cm] (P4){AP
                    selects a set of STAs for \\UL Tx and removes those with
                    \\expired timer from polling list};
                \node[process, below of = P4, node distance = 1.4cm] (P5) {UL
                    OFDMA frame exchange\\ with selected STAs};
                \node[process, below of = P5, node distance = 1.2cm] (P6) {AP
                    resets timer for STAs\\that successfully transmit};
                \node[process] at (-2.2, -3) (P7) {Channel\\contention};

                \path[arrow_line](start)--(D1);
                \path[arrow_line](D1)--node{no}(P1);
                \path[arrow_line](P1)--(P2);
                \path[arrow_line](D1)--node{yes}(D2);
                \path[arrow_line](D2.east) -| node[anchor=north, near start]{UL}
                    (D4.north);
                \path[arrow_line](D2.west) -| node[anchor=north, near start]{DL}
                    (D3.north);
                \path[arrow_line] (P2.east) -- ++(0.3,0) -| ++(0,-8.0) -|
                    ++(-9, 0) |- (P7.west);
                \path[arrow_line] (D3.east) -- node[pos=0.5,above]{no} (D4);
                \path[arrow_line] (D3) -- node[pos=0.5]{yes}(P3);
                \path[arrow_line] (D4) -- node[pos=0.5]{no}(P4);
                \path[arrow_line] (D4.east) -- node[pos=0.5]{yes} ++(2.55, 0);
                \path[arrow_line] (P4) -- (P5);
                \path[arrow_line] (P5) -- (P6);
                \path[arrow_line] (P6) -- ++(0, -.8);
                \path[arrow_line] (P7) |- (D1.west);
                \path[arrow_line] (P3.west) -- ++(-1.55, 0);
            \end{tikzpicture}
        }
    \captionsetup{belowskip=-19.5pt}
    \caption{Flow chart of the algorithmic procedure at AP side}
    \label{fig:ap_flow_chat}
\end{figure}

\section{Performance Evaluation} \label{performance_evaluation}
We validate the performance of the proposed A2P algorithm using the network
simulator, NS-3. It is a system-level discrete event network simulator
written in C++ that allows 
simulating the functions of the various elements of the network
stack~\cite{ns3, 2_Magrin}.

We compare A2P to the following baseline schemes:
\begin{enumerate*}
\item \emph{EDCA}: Disabling OFDMA to only use EDCA.
\item \emph{OFDMA}: Utilising only OFDMA, where all STAs in the
    network are polled. EDCA is disabled using MU EDCA Parameters Set. The MU
    EDCA timer is set to the maximum allowable duration of \qty{2088.96}{\ms},
    thus EDCA can still be used sporadically, e.g., for the initial packet of a
    new data flow.
\item \emph{OFDMA + EDCA}: Utilising both OFDMA and EDCA concurrently. In this
    case, the MU EDCA Parameters Set is not used.
\end{enumerate*}

The evaluation is based on the following performance indicators:
\begin{enumerate*}
    \item \emph{End-to-End (E2E) Delay}: The total round-trip time from the
        UL packet generation at a STA to the DL packet reception by
        this STA.
    \item\emph{Packet Loss Ratio}: Ratio of lost and outdated (i.e., arriving
        outside the TX delay budget)
        packets to the total number of transmitted packets.
    \item \emph{Wake Up Delay}: Delay of the first packet announcing an
        activated microphone.
\end{enumerate*}
\subsection{Simulation Setup}
\begin{table}[tb]
    \captionsetup{belowskip=3pt}
    \vspace{2pt}
    \caption{List of simulation parameters}
    \vspace{-0.3cm}
    \label{table:sim_para}
    \begin{center}
        {\renewcommand{\arraystretch}{1.5}%
        \begin{tabular}{l c}
            \hline
            \textbf{Parameter} & \textbf{Value} \\
            \hline
            Carrier frequency & \qty{5}{\giga\hertz} \\
            Bandwidth & \qty{40}{\mega\hertz} \\
            Guard Interval & \qty{0.8}{\micro\second} \\
            Propagation Loss Model & IEEE Indoor Model B \cite{indoorModel}\\
            MCS Index & \num{8} \\
            Resource Unit Type & 26\hyphen tone only\\
            Transmit Opportunity & \qty{2080}{\micro\second} \\
            AP Access Request Interval (ARI), $T$ & \qty{16}{\micro\second} \\
            MU EDCA Timer & \qty{40}{\milli\second} \\
            EDCA Access Category & VO \\
            Downlink Payload Size & \qty{500}{\byte} \\
            Uplink Payload Size & \qty{740}{\byte} \\
            Inter-Packet Interval, $X$ & \qty{5}{\milli\second} \\
            UL Packet Generation Window Part, $B$ & \qty{1}{\milli\second}
            \vspace{0.2cm}\\
            Off- to on- \& On- to off-time interval, $\tau$  &
            \pbox{10cm}{Bounded Exp.
            Distribution, \\(mean \qty{10}{\second}, max \qty{25}{\second})}
            \vspace{0.2cm} \\
            Number of STAs & \num{100} \\
            Duration of Simulation & \qty{30}{\second} \\
            \hline
        \end{tabular} }
    \end{center}
    \vspace{-0.8cm}
\end{table}

In the simulation, we  model a single basic service set (BSS) comprising
multiple STAs and a single AP that support the 802.11ax standard. The traffic
parameters are based on those of a state-of-the-art teleconferencing
application.
Each STA generates constant bit rate UDP traffic of approximately
\qty{1.2}{Mbps} (each packet contains $M = 240$ samples with $N = 24$ bit
resolution, and $W = 20$ byte application-layer header is used). The AP
broadcasts UDP traffic to all STAs at a rate of around \qty{0.1}{Mbps} (each
packet contains $P = 240$ samples with $Q = 16$ bit resolution, and the same
header size $W = 20$ byte is used). All traffic is set to the Voice (VO) AC.

We ensure that the transmission power is sufficiently high that the probability
of packet loss due to channel errors is negligibly small. Hence, in the
considered setup the packets can be lost only due to collisions or delays.
Furthermore, the impact on performance of STAs implicitly reporting their buffer
status is negligible, and thus disabled.

Not all STAs transmit UL data in any given experiment; those transmitting are
called \emph{active}, while those that do not send any data are called
\emph{idle}. The total number of STAs associated with the AP is equal to
\num{100}. Prior to the start of simulation, two sets of STAs are randomly
selected; \emph{initial} \num{8} active STAs, and a varying number of
\emph{joining} STAs (that will become active during the experiment). The AP
establishes Block ACK agreements with each STA at the start of simulation (cf.,
system setup in Figures~\ref{fig:sta_flow_chat} and~\ref{fig:ap_flow_chat}),
thus determining the ACK sequence that will be employed throughout the
simulation. The initial \num{8} STAs randomly transmit their first packets
within $B = 1$~\unit{\milli\second} after the system setup. After this step, the
joining STAs randomly decide when to start transmitting (referred to as
\emph{on-time}) and when to stop (i.e., \emph{off-time}). The time between the
consecutive UL data streams on the same STA (between off-time and next
on-time) and the stream duration (time between on-time and off-time), $\tau$, follow an exponential distribution with a mean of \qty{10}{\second}
and an upper bound of $\qty{25}{\second}$. This on-off transmission pattern of the joining STAs
continues for the duration of the simulation. This simulation scenario depicts a
teleconferencing system where participants wish to join or leave the ongoing
discussion. Table~\ref{table:sim_para} lists the parameter values used in the
experiments.
\vspace{-0.2cm}
\subsection{Discussion of Results}
\begin{figure}[tb]
    \centering
    \vspace{2pt}
    \includegraphics[scale=.45]{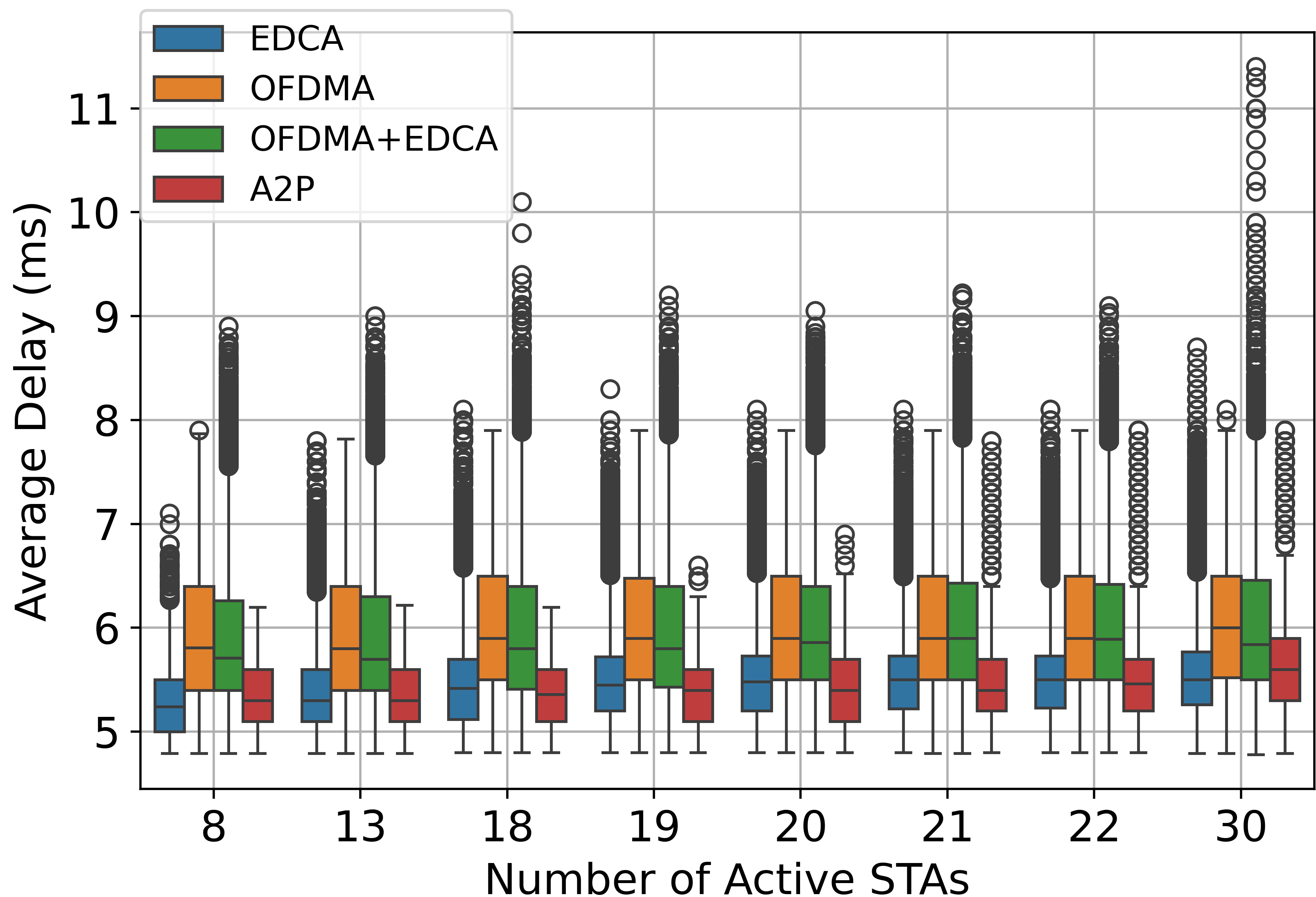}
    \captionsetup{belowskip=-5pt}
    \caption{Performance evaluation: average delay}
    \label{fig:all_algo_delay}
\end{figure}
\begin{figure}[tb]
    \centering
    \includegraphics[scale=.48]{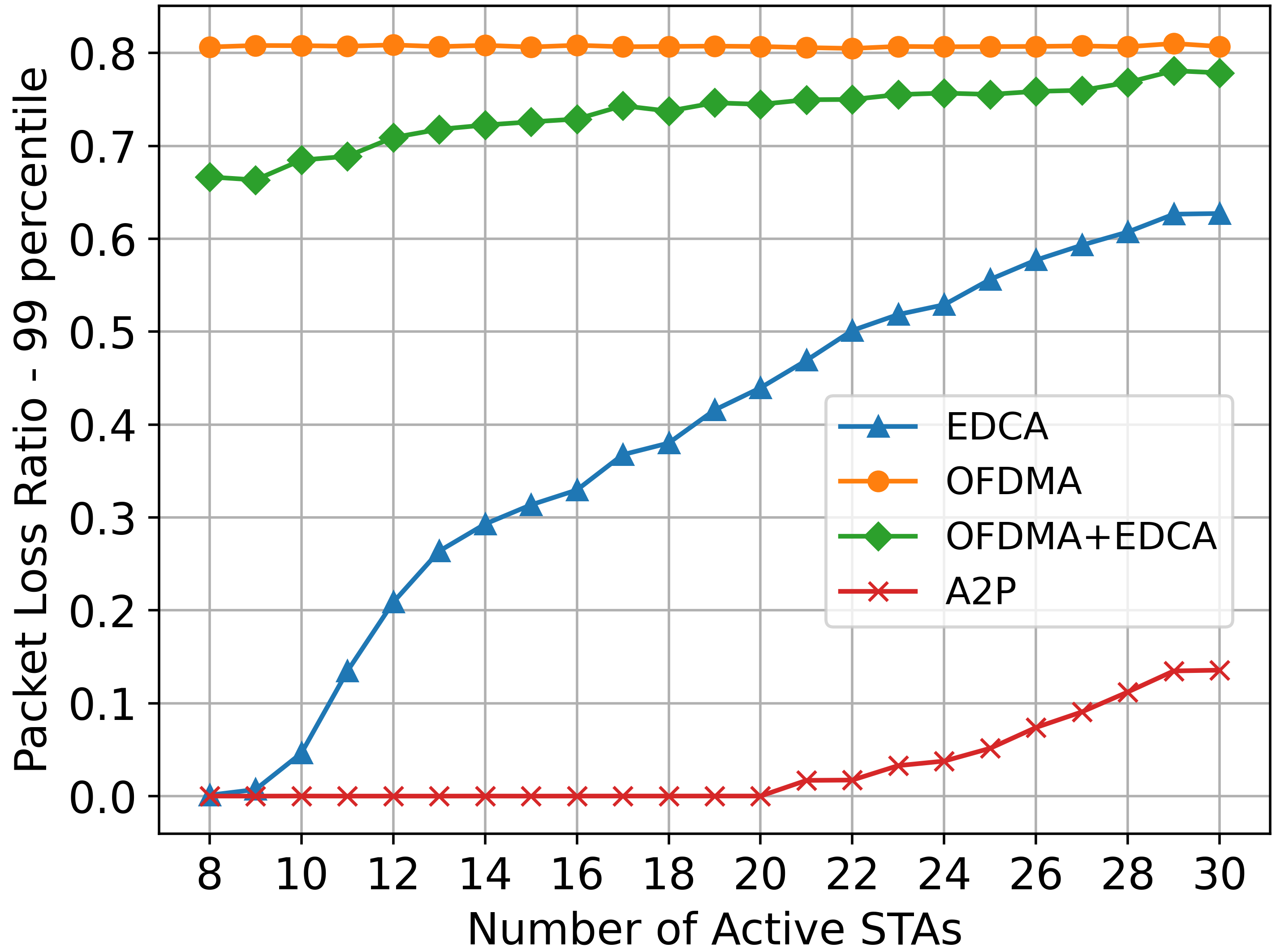}
    \captionsetup{belowskip=-15.9pt}
    \caption{Performance evaluation: 99\% packet loss ratio}
    \label{fig:all_algo_pktLossRatio}
\end{figure}
\begin{figure}[t]
    \centering
    \vspace{5pt}
    \includegraphics[scale=.45]{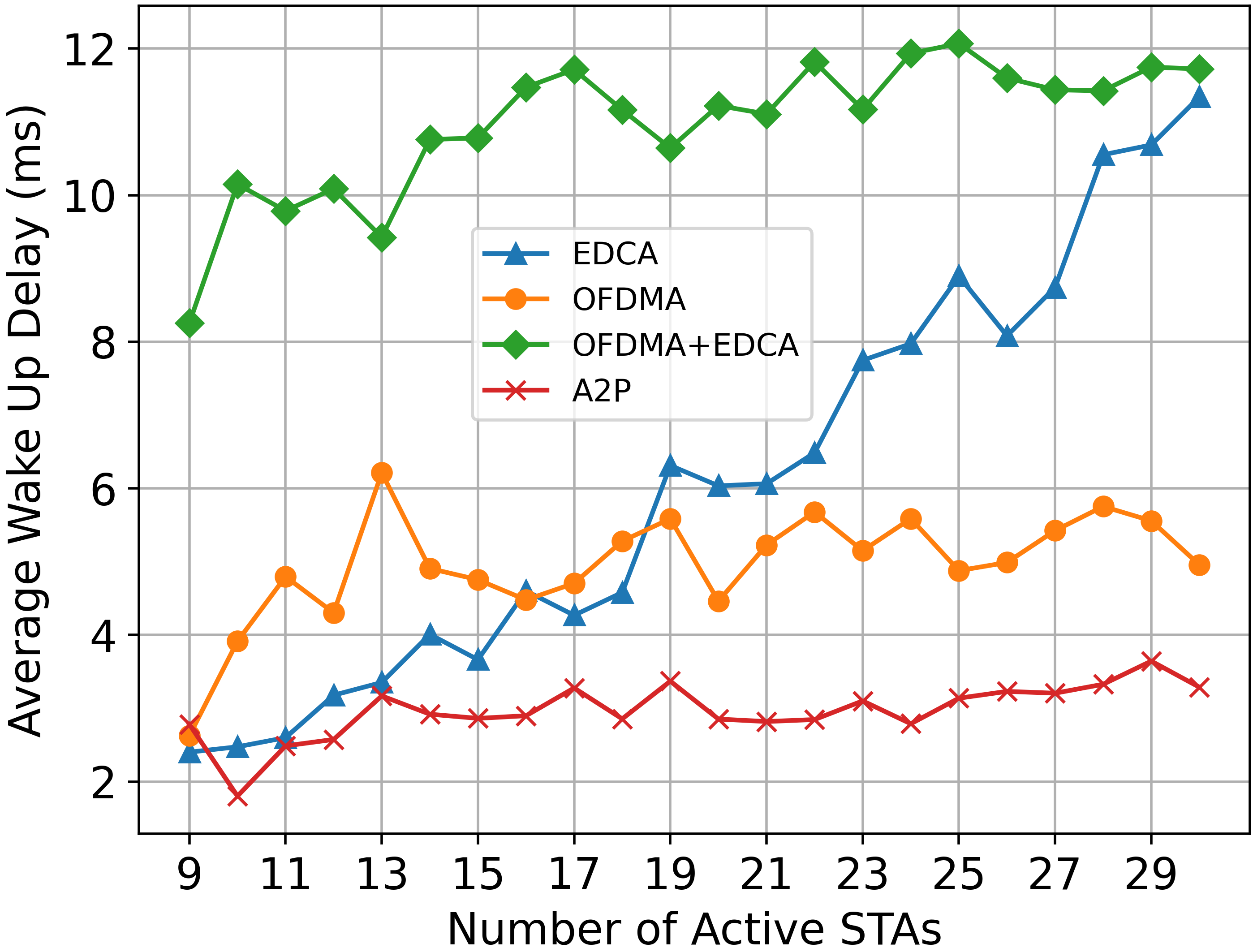}
    \captionsetup{belowskip=-15pt}
    \caption{Performance evaluation: average wake up delay}
    \label{fig:all_algo_wakeupTime}
\end{figure}
In this section, we compare the performance of the four schemes in terms of the
main metrics for the considered application, namely delay and packet loss ratio.
Figure~\ref{fig:all_algo_delay} represents the round-trip delay (from microphone
to headphones)  for different numbers of active STAs. Here by active STAs we
mean the number of joining STAs plus the initial \num{8} STAs. We show the
results as a box plot, where the median, lower and upper quartiles and extreme
values of delay can be clearly seen. A2P and EDCA both show relatively low
median and interquartile range compared to OFDMA and OFDMA+EDCA. However,
for EDCA the height of outliers increases significantly when moving from 8 to 19
active STAs.
With an UL delay budget of \qty{5}{\ms} and no budget constraints on DL
broadcasts, outliers are primarily attributed to DL transmissions. While EDCA DL
transmissions compete for the channel with UL transmissions and are vulnerable
to losses due to collisions, A2P manages to orchestrate UL and DL transmissions
without competition. The higher delay experienced by OFDMA and OFDMA+EDCA is due to additional latency of polling the large number of idle
STAs.

Besides the delay, we also compare the schemes in terms of the system
capacity, i.e., the number of STAs that we can support without experiencing
packet loss. Figure~\ref{fig:all_algo_pktLossRatio} shows that, in one BSS, up
to 20 and 27 devices are supported by A2P with \qty{0}{\percent} and
\qty{10}{\percent} packet loss, respectively. With similar QoS requirements,
EDCA supports a maximum of 10 active STAs. Furthermore, the poor performance of
OFDMA and OFDMA+EDCA schemes utilising legacy polling algorithm is caused by
inability to poll the STAs with UL data on time.

With EDCA, the average latency of initial packets announcing a
STA's activation (wake up delay) drastically increases with the growing number of active STAs due to contention. 
Nonetheless, as shown in Figure~\ref{fig:all_algo_wakeupTime}, A2P
addresses this by disabling EDCA on transmitting STAs, thereby reducing the
number of contending STAs.  In particular, the average wake up delay is less than
\qty{4}{\ms} for up to \num{30} active STAs. 
When an inactive device with disabled EDCA is reactivated after the MU EDCA
timer expires, the initial packet is transmitted through EDCA, thereby reducing
the wake up delay compared to an UL OFDMA round-robin polling scheme.
Note that OFDMA provides relatively low delay even in case of high number of
active STAs, because the activated STA still uses EDCA for the initial packet
due to expiration of the MU EDCA timer. In the OFDMA + EDCA case, OFDMA and EDCA operate independently, thus the performance is degraded due to
cumulative effect of the drawbacks of each scheme. More specifically, the
channel time is wasted on both EDCA contention between the active STAs and on
the transmission of unnecessary trigger frames to poll idle STAs.

\section{Conclusion} \label{conclusion}
In this work, we proposed the A2P polling algorithm that empowers Wi-Fi to meet
the stringent QoS requirements of real-time wireless systems.
Our approach leverages the combination of two channel access schemes, EDCA and
OFDMA, where OFDMA is used to orchestrate transmissions of active stations,
while EDCA provides a faster way of indicating the active state of a station.
More specifically, the AP maintains the list of active stations that are
regularly polled to update the information about their buffer statuses. Using a
representative use case of a dense teleconferencing system comprising a large
number of stations, we prove with simulation that A2P outperforms the legacy
schemes, such as EDCA only, OFDMA only and OFDMA+EDCA.
In particular, with 40 MHz bandwidth A2P supports up to 20 stations without packet loss, and achieves the lowest delay.

We plan to use mathematical modelling to develop algorithms
for the optimal selection of parameters (e.g., ARI) that provides efficient
polling and data transmission.
\section*{ACKNOWLEDGEMENTS}
This research was funded by the ICON project VELOCe (VErifiable, LOw-latency
audio Communication), realized in collaboration with imec, with project support
from VLAIO (Flanders Innovation and Entrepreneurship). Project partners are
imec, E-Bo Enterprises, Televic Conference, and Qorvo.

\printbibliography[title={References}]
\end{document}